\documentclass{aa}  

\usepackage{graphicx}
\usepackage{bm}
\usepackage[markup=draft,commentmarkup=uwave]{changes}
\usepackage[varg]{txfonts}
\usepackage[pdftex,
            colorlinks = true,
            linkcolor = blue,
            urlcolor  = blue,
            citecolor = blue,
            anchorcolor = blue]{hyperref}
\newcommand{\psp}{\emph{PSP}}
\renewcommand{\sc}{\rm sc}

\renewcommand{\vec}[1]{\mathbf{#1}}

\newcommand{\aave}[1]{\left\langle #1\right\rangle}

\newcommand{\Va}{V_{\rm A}}

\newcommand{\Vsw}{V_{\rm SW}}
\newcommand{\rms}{\rm rms}

\newcommand{\e}[1]{\times 10^{#1}}

\definechangesauthor[name=JCP, color=violet]{JCP}
\begin{document}

    \title{Applicability of Taylor's Hypothesis during Parker Solar Probe perihelia}
    %\title{Disentangling spacetime variations in Parker Solar Probe solar wind turbulence observations}
   %\subtitle{I. Overviewing the $\kappa$-mechanism}

   \author{Jean C. Perez
          \inst{1}\fnmsep\thanks{Corresponding author: Jean C. Perez, jcperez@fit.edu}
          \and
          Sofiane Bourouaine\inst{1,2}
          \and
          Christopher H.~K.~Chen\inst 3
          \and
          Nour E. Raouafi\inst 2
          }

   \institute{Department of Aerospace, Physics and Space Sciences, Florida Institute of Technology, 150 W. University blvd, Melbourne, Florida, 32901, United States of America.\\
              \email{jcperez@fit.edu}
         \and
             Johns Hopkins University Applied Physics Laboratory, Laurel, MD, United States of America.\\
             \email{sofiane.bourouaine@jhuapl.edu,nour.raouafi@jhuapl.edu}
        \and
        School of Physics and Astronomy, Queen Mary University of London, London E1 4NS, UK\\
        \email{christopher.chen@qmul.ac.uk}
\\
%             \email{c.ptolemy@hipparch.uheaven.space}
             }

   \date{Received XXXX, 2020; accepted YYYY, 2020}

    \abstract
    % context heading (optional) %leave it empty if necessary 
    {We investigate the validity of Taylor's Hypothesis (TH) in the analysis of Alfv\'enic fluctuations of velocity and magnetic fields in solar wind streams measured by \emph{Parker Solar Probe} (\psp)~during the first four encounters. The analysis is based on a recent model of the spacetime correlation of Magnetohydrodynamic (MHD) turbulence~\citep{bourouaine19}, which has been validated in high-resolution numerical simulations of strong Reduced MHD (RMHD) turbulence~\citep{perez20b}. We use \psp~velocity and magnetic field measurements from 24~h intervals selected from each of the first four encounters. The applicability of TH is investigated by measuring the parameter $\epsilon=\delta u_0/\sqrt{2}V_\perp$, which quantifies the ratio between the typical speed of large-scale fluctuations, $\delta u_0$, and the local perpendicular \psp~speed in the solar wind frame, $V_\perp$. TH is expected to be applicable for $\epsilon\lesssim0.5$ when \psp~is moving nearly perpendicular to the local magnetic field in the plasma frame, irrespective of the Alfv\'en Mach number $M_{\rm A}=\Vsw/\Va$, where $\Vsw$ and $\Va$ are the local solar wind and Alfv\'en speed, respectively. For the four selected solar wind intervals we find that between 10\% to 60\% of the time the parameter $\epsilon$ is below 0.2 when the sampling angle (between the spacecraft velocity in the plasma frame and the local magnetic field) is greater than $30^\circ$. For angles above $30^\circ$, the sampling direction is sufficiently oblique to allow one to reconstruct the reduced energy spectrum $E(k_\perp)$ of magnetic fluctuations from its measured frequency spectra. The spectral indices determined from power-law fits of the measured frequency spectrum accurately represent the spectral indices associated with the underlying spatial spectrum of turbulent fluctuations in the plasma frame. 
    %{\color{red}: I would also mention that TH has to be used with cautious as $\omega=k_\perp V_\perp$ is not accurate due to broadening that occurs because of the angle of $\vec k_\perp$ in the perpendicular $\vec k_\perp$ plane. Therefore, $\omega=\vec k_\perp \vec V_{\perp}$ is the correct one} 
    Aside from a frequency broadening due to large-scale sweeping that requires careful consideration, the spatial spectrum can be recovered to obtain the distribution of fluctuation's energy  among scales in the plasma frame. }
    %Conclusions
    % {}


   \maketitle
%
%-------------------------------------------------------------------

\section{Introduction}

The analysis of spacecraft signals invariably requires a number of assumptions to properly interpret temporal variations in terms of their corresponding spacetime variations in the plasma frame of reference, defined as the frame where the mean plasma bulk velocity is zero. The most common assumption used in the analysis of turbulent signals far from the Sun is the well known Taylor's Hypothesis (TH)~\citep{taylor38}, which posits that the temporal variation of spacecraft signals simply arises from the advection of ``frozen'' structures by the measuring probe. Although TH is almost universally assumed, implicitly or explicitly, in most analyses of solar wind observations~\citep{bruno13,chen16}, its accuracy and applicability to the interpretation of spacecraft observations is still not completely understood~\citep{verscharen19,narita17a}. 

The use of TH in solar wind observations is often justified on the simple assumption that the relevant characteristic speeds associated with linear and nonlinear processes in the plasma frame, such as the typical root-mean-squared (\rms) speed $\delta u_0$ at the injection scale and Alfv\'en speed $\Va$, are much smaller that the solar wind speed $\Vsw$ \citep{matthaeus82,perri10}. When $\Vsw\gg\delta u_0,\Va$, the plasma dynamics is assumed to be ``frozen'' in the plasma frame, and therefore standard correlation and spectral analysis of time signals is directly interpreted as spatial analysis, where the time $t$ can be associated with spatial coordinate $s=-\Vsw t$ in the ``upstream'' direction ($-\vec\Vsw$), which near 1~au is mostly anti-radial\footnote{Note that the negative sign comes from the fact that in reality $s=Vt$, where $V$ is the spacecraft velocity in the plasma frame, which near Earth is $V=-\Vsw$}. This relationship between space and time also implies a relationship between frequency and wavevector, $\omega=\vec k\cdot\vec\Vsw$, commonly used in the interpretation of frequency spectra of turbulent fluctuations in the solar wind. This frequency-wavevector relation is easily understood because when the magnetic field is approximately time-independent in the plasma frame, spacecraft frequencies $\omega=k_s\Vsw$ are mostly due to the Doppler-shift of zero plasma-frame frequencies. Here $k_{\rm s}$ is the ``streamwise'' component of the wave vector ($k_s$) in the plasma frame. 

As Parker Solar Probe (\psp)~\citep{fox16} reaches closer to the Sun, TH may lead to less accurate or even invalid results~\citep{klein14,bourouaine18}, and thus a new methodology is needed to interpret \psp~observations beyond TH. The expectation that TH may not be valid for \psp~measurements in the near-Sun solar wind has spurred a renewed interest in the fundamentals of the applicability of TH to solar wind observations and how the analysis of solar wind signals may differ for \psp~measurements~\citep{howes14,klein14,klein15,bourouaine18,narita17a,huang19,chhiber19,bourouaine19,perez20b}. A few of these works, which are based on specific assumptions that apply to Alfv\'enic turbulence, have suggested that under certain conditions, TH may still hold even when $\Vsw\sim\Va$~\citep{klein14,bourouaine19}. 

\cite{bourouaine19}, BP19 hereafter, proposed a new methodology to interpret turbulent signals beyond TH based on a recent model of the spacetime correlation of Magnetohydrodynamic (MHD) turbulence that was validated for strong MHD turbulence in high-resolution numerical simulations of Reduced MHD (RMHD) turbulence~\citep{perez20b}. This new methodology, which assumes that the turbulence is Alfv\'enic and highly anisotropic  ($k_\|/k_\perp\ll1$ where $k_\|$ and $k_\perp$ are the parallel and perpendicular components of the wavevector with respect to the magnetic field), depends on a single dimensionless parameter $\epsilon=\delta u_0/\sqrt2 V_\perp$ where  $\delta u_0$ is the rms value of the outer-scale fluid velocity (above the onset of the inertial range) and $V_\perp$ is the field-perpendicular velocity of the spacecraft in the plasma frame. TH is recovered in this model in the limit when $\epsilon\rightarrow0$, independent of the Alfv\'en Mach number $M_{\rm A}=\Vsw/\Va$.~\cite{bourouaine20} successfully applied this methodology to \emph{Helios} observations near 0.6~au and found that spectral power laws can be reliably measured as long as $\epsilon$ remains below 0.5. It is yet unknown if this relationship is applicable to \psp~observations near perihelia, which motivates the present work.

In this work we investigate the validity of TH in the first close encounters of \psp~in the framework of the BP19 methodology and evaluate the validity (and accuracy) of TH hypothesis by empirically estimating the dimensionless parameter $\epsilon$ for selected intervals during the first four encounters. This paper is organized as follows. In section~\ref{sec:BP19} we briefly summarize the BP19 model for the analysis of turbulent signals without assuming TH and discuss how it differs from recently related works.  In section~\ref{sec:data} we describe the \psp~data and methodology for the analysis of power spectral density of magnetic fluctuations in the context of BP19 model. In section~\ref{sec:results} we present the results of our analysis and in section~\ref{sec:conclusions} we conclude.
%We find that for the first four encounters TH can still be used to measure spectral power laws with reasonable accuracy when the spacecraft velocity in the plasma frame makes an angle of more than $30^\circ$ with respect to the local magnetic field. 
%{\color{red} we may also just remove this following sentence if you like}. For angles smaller than $30^\circ$, {\color{blue} a new analytical derivation using the BP19 model is needed and worth to be investigated, however, this is not the scope of this paper} .

% \section{Data Analysis and Methodology}\label{sec:data}

%\comment[id=JCP]{We need to expand to describe the data and add references}

% As mentioned in the previous section, a key feature of this model is that each turbulent field is separated into an outer-scale and a small scale component. 

\section{Analysis of turbulent measurements beyond TH\label{sec:BP19}}

For non-compressible and transverse, Alfv\'en-like, velocity $\delta\vec v$ and magnetic field $\delta\vec B$ fluctuations, Kraichnan's idealized sweeping model of Hydrodynamics~\citep{kraichnan65,wilczek12} was extended to strong MHD turbulence~\citep{bourouaine19,perez20b} to model the spacetime correlation function of Elsasser fields, $\vec z^\pm\equiv\delta\vec v\pm\delta\vec B/\sqrt{4\pi\rho}$, where $\rho$ is the plasma mass density. In this model, the spacetime correlation function is predominantly the result of sweeping of small scale fluctuations by large-scale ones, a hypothesis that was thoroughly validated against numerical simulations of Reduced MHD turbulence~\citep{perez20b}.  One key feature of this model is that fluctuating fields are split into outer-scale and small-scale fluctuations, i.e., it is assumed that
\begin{eqnarray}
\vec v=\vec v'+\delta\vec v,\quad\vec B=\vec B'+\delta\vec B
\end{eqnarray}
where primed variables, such as $\vec v'$ and $\vec B'$, are considered to be random variables describing eddies in the energy containing range (or outer scale) with known probability distribution functions, and $\delta\vec v,~\delta\vec B$ represent fluctuations at smaller scales. The role of the outer-scale velocity is to produce random advection (sweeping) of small-scale structures, while the role of outer-scale magnetic field is to randomly modify the background to provide a ``local magnetic field'' along which small-scale fluctuations propagate, which defines the field-parallel direction. An important question that one may ask is how far below the inertial range is Kraichnan's sweeping hypothesis valid. Due to the phenomenological nature of the sweeping models, both for HD and MHD, a quantitative answer is not possible. However, numerical simulations have validated the sweeping effect in HD~\citep{he06,verma20} as well as in MHD~\citep{perez20b}. In the latter case, the sweeping effect is observed to be present at scales that are approximately below one quarter of the outer scale, defined at the onset of the inertial range. 
Lastly, it is worth mentioning that spacetime correlations and the turbulence decorrelation time have been investigated in the context of the MHD turbulence by a number of authors~\citep{zhou10,matthaeus10,matthaeus16,servidio11,narita13,narita17a,weygand13} and recently in the framework of weak MHD turbulence~\citep{perez20a}. The main difference between the model of the spacetime correlation in the works of ~\cite{bourouaine19,perez20b} and these previous works is that the sweeping effect is purely hydrodynamic.

The relation between the spacecraft frequency spectrum and the three-dimensional power spectrum in the plasma frame that follows from this ``sweeping'' model has the form
\begin{equation}
        P^\pm_{\sc}(\omega) =\int \frac{P^\pm(\vec k_\perp,k_\|)}{\epsilon k_\perp V }g\left(\frac{\omega+\vec k_\perp\cdot\vec V_\perp+k_\|V_\|}{\epsilon k_\perp V}\right)d^2k_\perp dk_\|,\label{eq:Gamma} 
\end{equation}
where $\epsilon\equiv \delta u_0/\sqrt 2V$, $\delta u_0$ is the \rms~value of the velocity ($\vec v'$) of the energy-containing eddies, and $g(x)$ is the probability density distribution of velocities in the energy-containing range along a given direction $\vec{\hat n}$, where $x\equiv\sqrt 2v'_n/\delta u_0$ denotes the velocity component ($v_n'$) normalized to its \rms~value $\delta u_0/\sqrt 2$. For the solar wind it is typically found that $g(x)$ is very close to Gaussian~\citep{bruno13}. The dimensionless quantity $\epsilon$ provides a convenient parameter to assess the validity of TH hypothesis, which is recovered the limit $\epsilon\rightarrow 0$ 
\begin{equation}
    \lim_{\epsilon\rightarrow0}\frac{1}{\epsilon k_\perp V }g\left(\frac{\omega+\vec k_\perp\cdot\vec V_\perp+k_\|V_\|}{\epsilon k_\perp V}\right)=\delta(\omega+\vec k_\perp\cdot\vec V_\perp+k_\|V_\|).\label{eq:THlimit}
\end{equation}
It is important to note that the validity of Eqn.~\eqref{eq:THlimit} does not require $M_{\rm A}\gg1$, as long as the turbulence is strongly anisotropic. Remarkably, the transformation kernel in Eqn.~\eqref{eq:Gamma} is found to be the same for both Elsasser fields $\vec z^\pm$, independent of cross-helicity. Fundamentally, the reason that this transformation is the same for both Elsasser fields is because it is determined entirely by sweeping from the same velocity field, $\vec v'$, of the energy-containing scales. Therefore, from simplicity we drop the labels ``$\pm$'' as the following analysis is valid for both Elsasser fields $\vec z^\pm$.

\begin{table*}
    \centering
    \begin{tabular}{ccccccc}
    \hline
        Encounter & Interval                               &   $r$ (au)  & Plasma Instrument \\  
        \hline\hline 
          E1  & 2018-11-05 15:30 to 2018-11-06 15:30  &   0.166          & SPC \\  
          E2  & 2019-04-04 16:00 to 2019-04-05 16:00  &   0.166          & SPC \\
          E3  & 2019-08-29 12:00 to 2019-08-30 12:00  &   0.191          & SPC  \\ 
          E4  & 2020-01-28 14:30 to 2020-01-29 14:30  &   0.131          & SPAN-ion\\
          \hline
    \end{tabular}
    \caption{Selected 24~h intervals, one from each of the first four encounters E1 to E4, used in our analysis. $r$ represents the average heliocentric distance for each interval. The last column indicates the primary instrument used in the analysis of plasma moments. With the exception of E3, all intervals are selected to be near \psp~perihelia.}
    \label{tab:intervals}
\end{table*}

This relation can be reduced to a simpler expression connecting spacecraft frequencies to the field-perpendicular wavevector $k_\perp$ by making the following assumptions: (1) The three-dimensional power spectrum $P=P(k_\perp,k_\|)$ is nearly isotropic in the field perpendicular plane, i.e., it does not depend on the orientation of $\vec k_\perp$; (2) the spectrum $P(k_\perp,k_\|)$ is strongly anisotropic with respect to the magnetic field direction, i.e., it is nearly zero for $k_\|\ll k_\perp$;  and (3) the spacecraft velocity in the plasma frame $\vec V$ is ``sufficiently oblique'', i.e., it satisfies $V_\perp/V_\|>>k_\|/k_\perp$. The first two assumptions are based on theoretical predictions from a number of phenomenological models of MHD turbulence~\citep{goldreich95,chandran08,boldyrev05,boldyrev06,perez09}, which have been verified in high-resolution numerical simulations~\citep{muller05,mason06,perez12}, and are expected to be present in solar wind observations~\citep{bieber96,saur99,horbury08,wicks10,chen11a,chen12}. The third assumption simply requires that the sampling angle $\theta_{VB}$, defined as the angle between the spacecraft velocity in the plasma frame and the magnetic field ($\tan\theta_{VB}\equiv V_\perp/V_\|$), be much larger than a critical angle $\theta_c$, determined by the anisotropy $\tan\theta_c\sim k_\|/k_\perp$, which is expected to be small for strongly anisotropic turbulence. However, because \psp~observations are single-point measurements an empirical determination of this critical angle is not straightforward. For simplicity, we provide an empirical estimate of this critical angle by assuming the turbulence is critically balanced~\citep{goldreich95}, i.e., the Alfv\'en propagation time is of the same order as the nonlinear energy-cascade time at each scale $l\sim 1/k_\perp$ in the inertial range, $k_\|V_A\sim k_\perp\delta v_l$, when the turbulence cascade is strong. In this critically-balanced state, the energy predominantly cascades to small perpendicular scales, resulting in a scale dependent anisotropy in which $k_\|/k_\perp$ becomes smaller at smaller scales. We thus estimate $k_\|/k_\perp\sim \delta u_0/\Va\equiv\tan\theta_c$ using $\delta u_0$ at the outer scale, which provides an overestimate of the critical angle. Under these three assumptions the relation between the frequency power spectrum as measured by the spacecraft and the reduced field-perpendicular spectrum $E(k_\perp)=\int_{-\infty}^\infty 2\pi k_\perp P(k_\perp,k_\|)dk_\|$ is~\citep{bourouaine19}
\begin{equation}
        P_{\sc}(\omega)=\int_0^\infty E(k_\perp)\frac{\bar g_\epsilon\left(\omega/k_\perp V_\perp\right)}{k_\perp V_\perp}dk_\perp\label{eq:pomega0}
\end{equation}
where
\begin{equation}
    \bar g_\epsilon(x)=\frac 2\pi\int_0^\pi\frac 1\epsilon g\left(\frac{x+\cos\phi}\epsilon\right)d\phi.
  \end{equation}
results from the integration over the angle $\phi$ in the scalar product $\vec k_\perp\cdot\vec V_\perp=k_\perp V_\perp\cos\phi$. Note that because we are neglecting the parallel spacecraft velocity, $V_\|$, we use $V_\perp$ to define $\epsilon=\delta u_0/\sqrt 2V_\perp$ instead of the spacecraft speed $V$. 
For a power law spectrum $E(k_\perp)=Ck_\perp^{-\alpha}$, using the change of variables $x=\omega/k_\perp V_\perp$ Eqn.~\eqref{eq:pomega0} becomes
\begin{equation}
    P_{\sc}(\omega)=\frac{\Lambda_{\alpha,\epsilon}}{V_\perp}E\left(\omega/{V_\perp}\right)\label{eq:pomega}
\end{equation}
where 
\begin{equation}
    \Lambda_{\alpha,\epsilon}\equiv\int_0^\infty f_{\alpha,\epsilon}(x)dx,\qquad f_{\alpha,\epsilon}(x)\equiv x^{\alpha-1}\bar g_\epsilon(x).\label{eq:lambda}
\end{equation}

Equation~\eqref{eq:pomega} shows that the frequency power spectrum exhibits the same power law of the underlying spatial energy spectrum $E(k_\perp)$, even when TH does not hold. Although a similar result was also found in~\citep{wilczek12,narita17a}, an important difference with BP19 is that the broadening parameter $\epsilon$ is controled by pure HD sweeping and therefore the scaling factor $\Lambda_{\alpha,\epsilon}$ is the same for both $\vec z^+$ and $\vec z^-$. It is worth mentioning that although the model was derived for Elsasser fluctuations, it can be extended to velocity and magnetic field fluctuations.

The scaling factor $\Lambda_{\alpha,\epsilon}$ can be calculated once empirical values of $\alpha$ and $\epsilon$ are determined. Therefore, the analysis of turbulent power laws from spacecraft measurements in this framework requires the accurate estimation of $\alpha$, $\delta u_0$ and $V_\perp$.  \cite{bourouaine20} applied this methodology to a three day interval at 0.6~au from \emph{Helios} measurements, and found that for the observed values of $\epsilon\lesssim 0.1$, the empirical value for $\Lambda_{\alpha,\epsilon}$ remained close enough to the value expected when TH is valid, approximately $\Lambda_{\rm TH}\simeq 0.7628$ for $\alpha=3/2$ and $\Lambda_{\rm TH}\simeq 0.7132$ for $\alpha=5/3$. In this case TH can still be applied to reconstruct the reduced perpendicular energy spectrum, as long as the sampling angle ($\theta_{VB}$) is much greater than $20^\circ$. If the spacecraft velocity is below this critical angle, which defines the acceptable ``obliqueness'' of the spacecraft, a different analysis that involves the field-parallel components of the wavevector and the spacecraft velocity is required. 
%\cite{bourouaine20} also found that TH can still remain a good approximation for values of $\epsilon$ up to 0.5. In the following analysis, we determine to what extent TH can be applied within this framework to the first four \psp~close encounters.

\section{Data Description and Methodology}\label{sec:data}
\begin{figure*}
    \centering
    \includegraphics{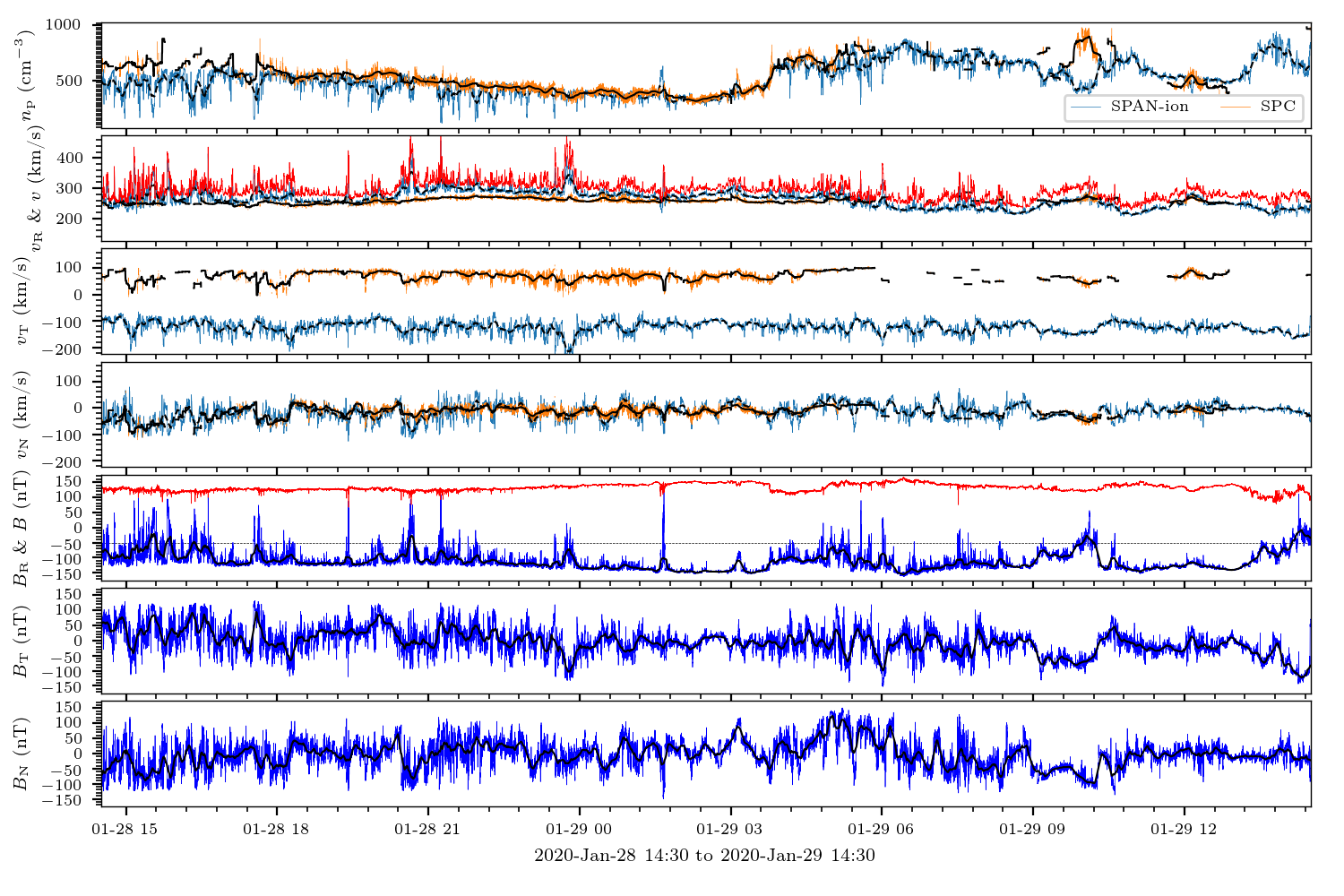}
    \caption{A 24~h PSP interval between 2020-Jan-28 at 14:30 and 2020-Jan-29 at 14:30. Top panel shows the proton number density as measured by both SPC and SPAN-ion. The second to forth panels (from top to bottom) show the bulk plasma velocity in the radial (R), tangential (T) and Normal (N) directions, respectively. The three bottom panels show the R, T and N components of the magnetic field. The red line in the second and fifth panels represents the solar wind speed and the magnitude of the magnetic field, respectively. During encounter E4, due to the high \psp's high speed at perihelion, a larger fraction of solar wind particles fall under SPAN-ion's than on SPC's field-of-view (FOV).}%, as can be seen in the first three panels.}
    \label{fig:e2}
\end{figure*}

%\subsection{Data description}
We use \psp~velocity and magnetic field measurements from a set of 24~hour intervals, shown in Table~\ref{tab:intervals}, during the first four close encounters covering heliocentric distances between $0.13$~au to $0.19$~au to test the validity of TH near \psp~perihelia. 
%The first encounter (E1) is taken as the period from XXX to YYY, the second encounter between \ldots.  
Proton number density and velocity are obtained from the moments of the velocity distribution functions measured by Solar Probe Cup (SPC) and the SPAN-ion on board the SWEAP instrument suite~\citep{kasper16}. The choice between SPC and SPAN-ion signals is made based on which instrument has the best Field-of-View (FOV) for each interval we analyze (see Table~\ref{tab:intervals}), while for those cases where it is not clear which instrument provides a better estimate of the moments, our analysis is performed with both signals to determine the sensitivity of our analysis to discrepancies between SPC and SPAN measurements. Magnetic field measurements are obtained from the fluxgate magnetometer (MAG) on board the FIELDS instrument suite~\citep{bale16}.  Velocity and magnetic field measurements, which are sampled with an average resolution of $0.874$~s and $0.22$~s, respectively, are re-sampled on the same temporal grid by averaging over a~1~s window. Figure~\ref{fig:e2} shows time signals of proton number density, radial and tangential velocity from SPC and SPAN-ion measurements, and magnetic field measurements during a day-long interval a few days before the fourth perihelion.
 
%\subsection{Data analysis}
The velocity and magnetic fields in the energy-containing range are obtained by performing the following moving averages over a time window $T$ at each time $t$
\begin{eqnarray}
    \vec v'(t)=\frac 1T\int_{t-T/2}^{t+T/2}\vec v(t')dt',\;\;\;\hbox{and}\;\;\; \vec B'(t)=\frac 1T\int_{t-T/2}^{t+T/2}\vec B(t')dt'\label{eq:ma}
\end{eqnarray}
where $T$ is chosen to correspond to the turbulence outer scale, which we define as the inverse of the spectral break frequency that separates the $f^{-1}$ range from the inertial range. In other words, we take $T=1/f_{\rm b}$ where $f_{\rm b}$ is the frequency at the onset of the inertial range. In the present analysis we take $T=8$~min, consistent with spectral break frequency estimates $f_{\rm b}\approx 2\e{-3}$~Hz from the first two encounters~\citep{chen20,parashar20}. Similar values for the spectral break frequency were also found by~\cite{bourouaine20b} when the power spectra were calculated within and outside the so-called Switchbacks (SB)~\citep{bale19,kasper19},  although larger values of $f_{\rm b}\approx 2\e{-2}$ have been reported outside SB regions~\citep{dudokdewit20}. The moving averages defined in Eqn~\eqref{eq:ma}, which act as a low-pass filter that removes fluctuations below the timescale $T$,  lead to smoother random time signals plotted as black lines in Figure~\ref{fig:e2}, representing the large-scale component of the corresponding quantities in the interval E4 described in Table~\ref{tab:intervals}. In the figure dash lines correspond to SPAN-ion measurements and solid lines correspond to SPC.

Once the time signals for the outer scale velocity fluctuations are obtained, their mean and root-mean-square (rms) values 
\begin{equation}
    \vec\Vsw=\aave{\vec v'},\qquad \delta u_0^2=\aave{|\vec v'|^2}-|\vec\Vsw|^2\label{eq:rms}
\end{equation}
are calculated, respectively. Here $\aave{\cdots}$ represents a suitable ensemble averaging procedure, which in practice is replaced by a temporal average under the assumption of ergodicity. The rms velocity $\delta u_0$ represents the root-mean-squared value of the bulk velocity associated with fluctuations that are larger than $T=8$~min, which can also be obtained from the total fluctuation energy between $f=0$ to $f=f_{\rm b}$, as calculated by~\cite{bourouaine20}. We are left to determine the spacecraft velocity in the plasma frame $\vec V$ and its angle with respect to the \emph{local} magnetic field $\theta_{VB}$
%{\color{red} the quantities below may should be all in prime? or may be not?}
%\comment[id=JCP]{I need to check $\vec V$ vs $\vec v'$ and see if it makes a difference, physically and/or in data}
%
\begin{equation}
    \vec V = \vec V_{\sc}-\vec\Vsw,\;\;\;\hbox{and}\;\;\;\theta_{VB}(t)=\arccos{\left(\vec{\hat V}\cdot\vec{\hat B'}(t)\right)}.\label{eq:thetavb}
\end{equation}
where $\vec{\hat V}$ and $\vec{\hat B'}$ represent unit vectors in the direction of $\vec V$ and $\vec B'$, respectively. Note that because the local magnetic field $\vec B'(t)$ is fluctuating, one obtains a distribution of angles corresponding to turbulent ``realizations'' at each time $t$. 

Equation~\eqref{eq:pomega} allows one to reconstruct the reduced energy spectrum $E(k_\perp)$ from the spacecraft frequency spectrum $P_{\sc}(\omega)$ as long as one considers intervals where the sampling angle is much larger than the critical angle $\tan\theta_c\approx \delta u_0/\Va$. Because $V_\perp$ (and therefore $\epsilon$) depends on the sampling angle $\theta_{VB}$, we classify measurements at each time $t$ into ten-degree angular bins $\Delta\theta=10^\circ$ centered around $\theta_i=10^\circ$ to $90^\circ$ in increments of $10^\circ$.

\begin{table*}[!h]
    \centering
    \begin{tabular}{ccccccccccc}
    \hline
        Encounter &  $V$  (km/s)  & $\delta u_0$ (km/s)  &  $\Va$ (km/s)  & $M_{\rm A}$ & $\theta_{\rm c}$ & $\chi_c$ (\%)  & $\chi_{30} (\%)$ & $\epsilon_{\rm c}$& $\epsilon_{30}$\\
        \hline\hline 
          E1       &  315     &  40      & 82  &  3.8  & $23^\circ$  & 56 & 43 & 0.24 & 0.19  \\
          E2       &  400     &  37      & 182 &  2.2  & $12^\circ$  & 63 & 10 & 0.34 &0.14\\
          E3       &  490     &  49      & 157 &  3.1  & $17^\circ$  & 52 & 18 & 0.25 & 0.15\\ 
          E4       &  300     &  38      & 114 &  2.6  & $22^\circ$  & 92 & 57 & 0.19 & 0.14 \\
          \hline
    \end{tabular}
    \caption{Relevant empirical parameters for the four 24~h intervals in Table~\ref{tab:intervals} used in our analysis. $V$ is the average spacecraft speed in the solar wind frame, $\delta u_0$ the rms value of velocity at the large scale ($T=8$~min), $\theta_{\rm c}$ is the critical angle above which the spacecraft can be considered nearly perpendicular to the local field, $\chi_{\rm c}$ is the fraction of the interval in which the sampling angle is above the critical angle $\theta_{\rm c}$ and $\chi_{30}$ is the fraction of the interval in which the sampling angle is above $30^\circ$.}
    \label{tab:results_all}
\end{table*}

The temporal autocovariance of magnetic field signals\footnote{Here we convert magnetic field signals to fluctuating Alfv\'en velocity using the local average plasma density.} $\vec b(t)\equiv\vec B(t)/\sqrt{4\pi\rho}$ is calculated for each sampling angle range using  conditioned correlation functions~\citep{bourouaine20,bourouaine20b}
\begin{equation}
    C(\tau,\theta_{VB}) = \aave{\vec b(t)\cdot\vec b(t+\tau)}_{\theta_{VB}}\label{eq:ctheta}
\end{equation}
where the ensemble average is calculated by averaging over those times $t$ for which the angle $\theta_{VB}$ fall within each of the angle bins defined above. From here, a frequency spectrum is obtained for each angle
\begin{equation}
    P_{\sc}(\omega,\theta_{VB})=\frac 1{2\pi}\int_{-\infty}^\infty C(\tau,\theta_{VB})e^{i\omega\tau}d\tau.\label{eq:ptheta}
\end{equation}
If the assumption of anisotropy and oblique sampling direction are satisfied, all $P_{\sc}$ spectra should result in the same spatial spectrum $E(k_\perp)$ independent of the sampling angle~\citep{bourouaine20}. It then follows from Equation~\eqref{eq:pomega}
% The resulting power spectral density $P_{\sc}(\omega)$ describes the fluctuation energy with frequencies between $\omega$ and $\omega+d\omega$, i.e.
% \begin{equation}
%     P_{\sc}(\omega)d\omega = E(k_\perp)dk_\perp.\label{eq:p_vs_e}
% \end{equation}
%For power law spectra, equation~\eqref{eq:pomega} can be used in the right hand side of~\eqref{eq:p_vs_e} to obtain
\begin{equation}
    E(\omega/\lambda V_\perp) = \lambda V_\perp P_{\rm sc}(\omega)\label{eq:eperp}
\end{equation}
where $\lambda\equiv\Lambda^{1/(\alpha-1)}$. This last expression shows that the spatial power spectrum can be reconstructed from the frequency power spectrum by using the re-scalings $k_\perp=\omega/\lambda V_\perp$ and $E=\lambda V_\perp P_{\sc}$. It is worth noting that although the expression $k_\perp=\omega/\lambda V_\perp$ provides a connection between frequency and wavenumber, it should not be interpreted in the same way as when TH applies. When TH is not valid, the fluctuation energy within a narrow frequency range around each spacecraft frequency $\omega$ cannot be associated with a narrow range of wavevectors around a single wavevector in the plasma frame. In the present case, the transformations $k_\perp=\omega/\lambda V_\perp$ and $E=\lambda V_\perp P_{\sc}$ simply provide a way to reconstruct the spatial power spectrum in terms of $k_\perp$ from the measured $P_{\sc}(\omega)$ at each angle $\theta_{VB}$.

\section{Results\label{sec:results}}
\subsection{Empirical evaluation of $\epsilon$ in the first four encounters}
Table~\ref{tab:results_all} shows the most relevant parameters, within the methodology described in the previous section, obtained empirically for all intervals considered in this work. For encounters E1 to E3 plasma measurements from the SPC instrument are used in the analysis, while plasma measurements from the SPAN-ion instrument are used in the analysis of encounter E4. 

The average spacecraft speed measured in the plasma frame ranges from $300$~km/s to nearly $500$~km/s, so that the interval selection covers both slow and fast solar wind streams. The rms of velocity at the outer scale ranges from $40$~km/s to $52$~km/s, and is much smaller than the spacecraft motion. The average Alfv\'en Mach number $M_{\rm A}$ is found between two and four, which means that under the usual assumptions, TH could be marginally applied at best, given that $M_{\rm A}$ is not much larger than one. However, because our analysis is based on the assumptions in the BP19 methodology, we do not require this as a condition. One condition that the BP19 methodology does require is that one can only consider intervals for which the sampling angle is greater that the critical angle $\tan\theta_{\rm c}\sim\delta u_0/\Va$ (assuming the turbulence is critically balanced), which we empirically find in the range from $10^\circ$ to $20^\circ$ for the four intervals we consider.

\begin{figure}
    \centering
    \includegraphics{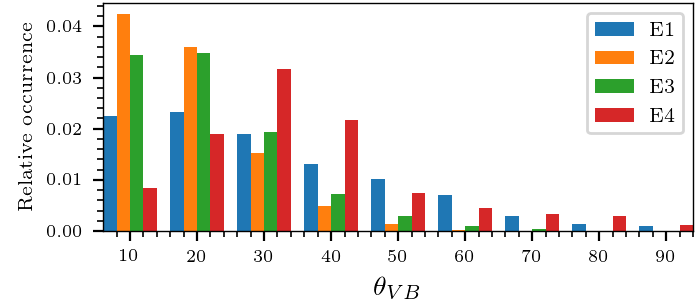}
    \caption{Normalized histograms showing the distribution of the sampling angle $\theta_{VB}$ defined in equation~\eqref{eq:thetavb} for selected 24~h intervals, one for each encounter from E1 to E4. The histograms indicate that for the selected intervals, small sampling angles have the highest occurrence, indicating that the spacecraft is often traveling nearly parallel to the local magnetic field. However, sampling angles above the critical angles shown in Table~\ref{tab:intervals} occur more than 50\% of the time. }
    \label{fig:epsilon_e2}
\end{figure}

Figure~\ref{fig:epsilon_e2} shows the distribution of sampling angles $\theta_{VB}$ resulting from our analysis of all four intervals. It is observed that for a substantial portion of the signal, the sampling direction is between $10^\circ$ and $20^\circ$, indicating that \psp~is very often flying nearly parallel to the local magnetic field.  For these small angles, the methodology described above does not apply requires further investigation. 

In order to quantify how often the spacecraft is sampling at a given angle, we define $\chi(\theta)$ as the fraction of the time in which the sampling angle is \emph{above} a certain value $\theta$
\begin{equation}
    \chi(\theta)\equiv \frac{\hbox{Number of samples where}~\theta_{VB}\ge\theta}{N_{\rm total}}
\end{equation}
where $N_{\rm tot}$ is the total number of samples. By a sample (or count) we mean an individual \psp~measurement out of the 86,400 records available in any 24~h interval at 1~s resolution. Table~\ref{tab:results_all} shows that the fraction of samples above the critical angle, $\chi_{\rm c}$, comprise 50\% to 90\% of the total count, allowing for at least one half of each interval for statistical analysis. However, we restrict the analysis for angles at or above $30^\circ$, for which the fractions $\chi_{30}$ shown in Table~\ref{tab:results_all}, is much lower, particularly worse for encounters E2 and E3. For the empirical values of $\delta u_0$ and $V$ shown in Table~\ref{tab:results_all} we can determine $\epsilon$ as a function of the sampling angle $\theta_{VB}$
\begin{equation}
    \epsilon=\frac{\delta u_0}{\sqrt 2V_\perp}=\frac{\delta u_0}{\sqrt 2V\sin\theta_{VB}}\label{eq:etheta}
\end{equation}
and obtain the values of $\epsilon$ at the critical angle ($\epsilon_{\rm c}$) and at the smallest angle we consider in the present analysis ($\epsilon_{30}$). We found that $\epsilon_{\rm c}\le0.35$ and $\epsilon_{30}\le 0.2$ across all intervals, which is  below the acceptable level of 0.5 obtained by~\cite{bourouaine20}.

\begin{figure*}
  \includegraphics{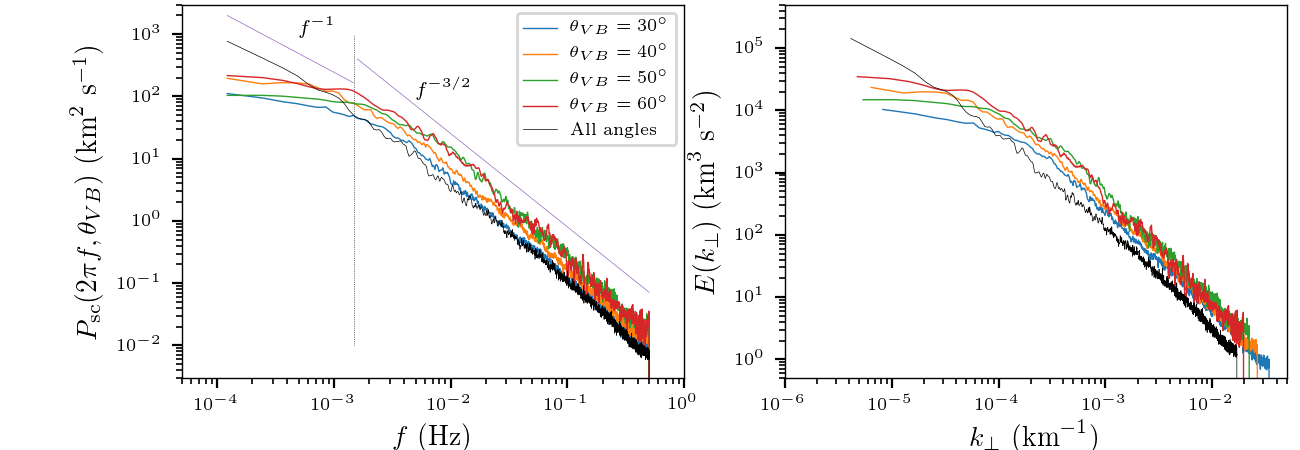}
  \caption{Left: Frequency spectrum of magnetic fluctuations $P_{\sc}(\omega,\theta_{VB})$ for angles $\theta_{VB}=30^\circ,~40^\circ,~50^\circ$ and $60^\circ$ corresponding to interval E4, as well as the full spectrum irrespective of the angle. The spectral break between the $1/f$ and $f^{-3/2}$ ranges for the unconditioned spectrum (black line) is found approximately (by inspection) at $f_{\rm b}\approx2\e{-3}$~Hz, indicated by the vertical dotted line.  Right: Spatial energy spectrum of magnetic fluctuations $E(k_\perp)$ reconstructed from the frequency spectra $P_{\sc}(\omega,\theta_{VB})$ for each sampling angle as well as by using the standard TH (black line). }\label{fig:pspectra}
\end{figure*}

\subsection{Reconstructing field-perpendicular spectrum $E(k_\perp)$}
In the following, we proceed to reconstruct the reduced energy spectrum $E(k_\perp)$ using data from within each angle bin above $30^\circ$ and for which we find a large statistical sample. For illustration purposes we concentrate on a 24~h interval just before the fourth perihelion (E4). Table~\ref{tab:results} shows the values of $\epsilon$, $V_\perp$ and the number of statistical samples (counts) associated with this interval for each ten-degree angle bin around $30^\circ,~40^\circ,~50^\circ$ and $60^\circ$.  In this statistical sample $\epsilon\le 0.2$, in which case the BP19 methodology in the TH limit ($\epsilon\rightarrow0$) is applicable, i.e., combining equations~\eqref{eq:Gamma} and~\eqref{eq:THlimit}
\begin{equation}
  P_{\sc}(\omega)=\int P(\vec k)\delta(\omega+\vec k\cdot\vec V)d^3k.\label{eq:pomega_TH}
\end{equation}
The most common form of TH follows from this expression by performing the integral in a cartesian coordinate system with one axis along the streamwise direction so that
\begin{equation}
  P_{sc}(\omega)=\frac 1V\int_{-\infty}^\infty E_{1D}(k_s)\delta(k_s+\omega/V) dk_s=\frac 1VE_{1D}(\omega/V).
\end{equation}
Here $E_{1D}(k_s)$ is the one-dimensional energy spectrum with respect to the streamwise (sampling) direction. This expression resembles equation~\eqref{eq:pomega} with $\Lambda_{\alpha,\epsilon}=1$ and leads to the familiar $\omega=-k_sV$ expression to relate spacecraft frequencies with streamwise wavenumber $k_s$.

\begin{table}
    \centering
    \begin{tabular}{cccccc}
    \hline
        $\theta_{VB}$ (degrees) & $\epsilon$  & $V_\perp$ (km/s) & Counts       & $\Lambda_{\alpha,\epsilon}$ \\
        \hline\hline
            30                  & 0.20        &  198             &   14305     &  0.7667  \\
            40                  & 0.15        &  254             &     6690    &  0.7649  \\
            50                  & 0.13        &  303             &   5548      &  0.7644   \\
            60                  & 0.11        &  342             &     3630    &  0.7639  \\
            \hline
    \end{tabular}
    \caption{Relevant parameter for each angle bin considered in our analysis with a sufficiently large number of samples. The parameter $\epsilon$ remains below 0.15, which allows for the application of the TH in the framework of BP19. For reference, at 1~s resolution the total number of counts in any given 24~h interval is 86400 samples. Also note that values for $\Lambda_{\alpha,\epsilon}$ are calculated for $\alpha=3/2$ and using equation~\eqref{eq:pscTH} $\Lambda_{\rm TH}\simeq 0.7628$.}
    \label{tab:results}
\end{table}

However, because we are interested in reconstructing the angle-integrated field-perpendicular spectrum $E(k_\perp)$, equation~\eqref{eq:pomega_TH} must be integrated in cylindrical coordinates. Using the change of variables $x=\omega/k_\perp V_\perp$ we obtain\footnote{Here it has been further assumed that $|k_\|V_\||\ll|\vec k_\perp\cdot\vec V_\perp|$}
\begin{equation}
  P_{\sc}(\omega) = \frac{\Lambda_{\rm TH}}{V_\perp}E(\omega/V_\perp)\label{eq:pscTH}%,\quad\hbox{where } \Lambda_{\rm TH} = \frac{\Gamma\left(\frac\alpha 2\right)}{\sqrt\pi\Gamma\left(\frac{\alpha+1}2\right)}
\end{equation}
where 
\begin{equation}
  \Lambda_{\rm TH} = \frac 2\pi\int_0^1 f_{\alpha,\rm TH}(x)dx,\quad f_{\alpha,\rm TH}(x)=\frac 2\pi\int_0^1\frac{x^{\alpha-1}}{\sqrt{1-x^2}}.%\frac{\Gamma\left(\frac\alpha 2\right)}{\sqrt\pi\Gamma\left(\frac{\alpha+1}2\right)}\label{eq:lambdaTH},
\end{equation}
Again, equation~\eqref{eq:pscTH} also resembles equation~\eqref{eq:pomega} with $\Lambda_{\alpha,\epsilon}$ replaced by $\Lambda_{\rm TH}$. In fact, it can be shown that $\Lambda_{\rm TH}=\lim_{\epsilon\rightarrow 0}\Lambda_{\alpha,\epsilon}$, consistent with the BP19 model in the limit when $\epsilon\rightarrow0$. The scaling factor $\Lambda_{\rm TH}$ is due to the integration with respect to the angle $\phi$ in the dot product $\vec k_\perp\cdot\vec V=k_\perp V_\perp\cos\phi$~\citep{bourouaine12,bourouaine13,martinovic19}.
%In addition to the scaling factor $\Lambda_{\rm TH}$ the angular integration in $\phi$ also leads to a non-negligible broadening of the function
%\begin{equation}
%  f_{\rm TH}(x)=\lim_{\epsilon\rightarrow0} f_{\alpha,\epsilon}(x)=\frac 2\pi\int_0^1\frac{x^{\alpha-1}}{\sqrt{1-x^2}}dx
%\end{equation}

%In light of this result, we reconstruct the reduced energy spectrum using TH according to equation~\eqref{eq:pscTH}.
Using equations~\eqref{eq:ctheta} and \eqref{eq:ptheta} within each statistical sample described in Table~\ref{tab:results}, $P_{\sc}(\omega,\theta_{VB})$ is calculated for magnetic field measurements $\vec b(t)$ according to equation~\eqref{eq:ptheta}. Left panel of Figure~\ref{fig:pspectra} shows frequency spectra $P_{\sc}(\omega,\theta_{VB})$ for the interval in encounter E4 for angles $30^\circ,~40^\circ,~50^\circ$ and $60^\circ$, as well as the full spectrum without imposing conditions on the angle (all angles). Note that the spectral break frequency $f_{\rm b}$ for the unconditioned spectrum (black line) is found (by inspection) approximately at $2\e{-3}$, consistent with our assumption that the outer scale corresponds to $T=8$~min. All five frequency spectra are consistent with a spectral index $\alpha=3/2$ in agreement with an independent analysis of the turbulence properties of this interval by~\cite{chen20b}. The right panel shows the reduced energy spectrum $E(k_\perp)$ reconstructed using equation~\eqref{eq:eperp} with $\lambda=\Lambda_{\rm TH}^{1/(\alpha-1)}=\Lambda_{\rm TH}^2\simeq 0.58$. The scaling factor $\Lambda_{\alpha,\epsilon}$ calculated for the empirical values of $\epsilon$ in each interval are found to differ from $\Lambda_{\rm TH}$ by less than 1\%, as shown in Table~\ref{tab:results}.

\begin{figure}
  \includegraphics{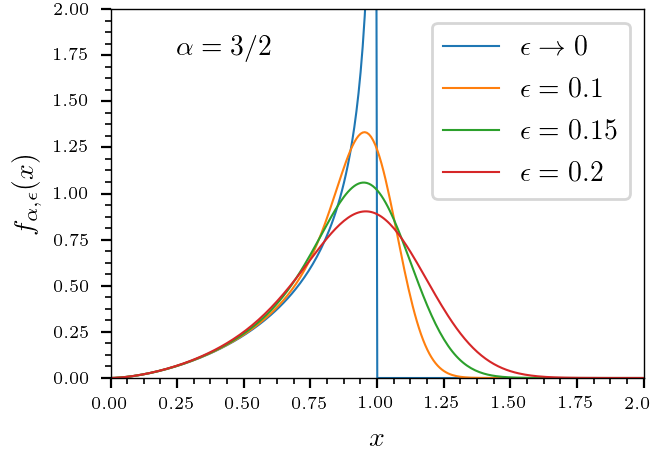}
  \caption{Function $f_{\alpha,\epsilon}(x)$ for $\alpha=3/2$ with $\epsilon$ values similar to those obtained empirically. It is observed that for $\epsilon\le 0.2$ the function peaks around $x=1$ corresponding to the TH limit, while the broadening around $x=1$ on the left side of the peak is similar for all $\epsilon$. The difference between using the BP19 phenomenology or TH is more pronounced on the right side of the peak, affecting small frequencies.}\label{fig:falpha_eps}
\end{figure}

One must however note that equation~\eqref{eq:eperp}, which follows from~\eqref{eq:pomega}, holds when the power law $E(k_\perp)$ extends from $k_\perp=0$ to $\infty$. In solar wind observations power-law behavior extends over a finite range from $k_{\min}$ to $k_{\max}$, and therefore the integration in equation~\eqref{eq:lambda} must be performed over the interval $[x_{\min},x_{\max}]$, where $x_{\min}=\omega/k_{\max}V_\perp$ and $x_{\max}=\omega/k_{\min}V_\perp$. In this case, the parameter $\Lambda_{\alpha,\epsilon}$ becomes a function of $\omega$, and $P_{\sc}(\omega)$ no longer exhibits the same power law as $E(k_\perp)$. However, because $f_{\alpha,\epsilon}$ is usually a localized function, it is possible to define an interval $[x_0,x_1]$ such that
\begin{equation}
  \int_{x_0}^{x_1}f_{\alpha,\epsilon}(x)dx\approx\int_0^{\infty}f_{\alpha,\epsilon}(x)dx= \Lambda_{\alpha,\epsilon},\label{eq:intf}
\end{equation}
in which case equation~\eqref{eq:eperp} remains approximately valid for frequencies in the range
\begin{equation}
  x_1k_{\min}V_\perp\le\omega\le x_0k_{\max}V_\perp.\label{eq:frange}
\end{equation}
Equation~\eqref{eq:frange} also reveals that the fluctuation energy measured in a narrow frequency bin $\delta\omega$ around $\omega$ arises from a broad range of wavenumbers inside the range $[k_{\min},k_{\max}]$, rather than a narrow wavenumber bin $\delta k$ around a streamwise wavenumber $k_s=\omega/V_\perp$, even in the TH limit when $\Lambda_{\alpha,\epsilon}\approx\Lambda_{\rm TH}$. For instance, Figure~\ref{fig:falpha_eps} shows that the function $f_{\alpha,\epsilon}(x)$, for values of $\epsilon$ similar to those obtained empirically and $\alpha=3/2$, becomes negligibly small above $x=2$. It can also be shown that its integral from $x_0=0.25$ to $x_1=1.25$ captures more than 90\% of $\Lambda_{\rm TH}$ for any value of $\epsilon\le 0.2$. In all cases the largest contribution to the fluctuation energy at a given frequency comes from $x\approx 1$, while the overall width of the function $f_{\alpha,\epsilon}(x)$ increases, somewhat asymmetrically, with increasing $\epsilon$. The broadening to the left of $x=1$, which affects high frequencies, remains largely unchanged with increasing $\epsilon$ when compared with the TH limit. On the right side of $x=1$, which affects low frequencies, the broadening is more significant for $\epsilon=0.2$. When putting everything together, equation~\eqref{eq:pscTH} approximately holds as long as
\begin{equation}
  0.2k_{\min}V_\perp\le\omega\le 1.25k_{\max}V_\perp,\label{eq:omg_range}
\end{equation}
assuming $E(k_\perp)$ exhibits a power law in the range $[k_{\min},k_{\max}]$. A possible consequence of the substantial broadening to the right of $x=1$ is that it could play a role in spectral breaks at low frequencies. However, because the model of the spacetime correlation holds for fluctuations with lengthscales that are much smaller than the outer scale and highly anisotropic, the validity of the BP19 phenomenology is less justified for frequencies (or wavenumbers) that are too close to the spectral break between the $1/f$ and $f^{-3/2}$ ranges, where the spectrum is more likely to be isotropic~\citep{wicks10,wicks11}. The validity of the BP19 model for the spacetime correlation has been verified in numerical simulations for inertial-range scales that are approximately below one quarter of the outer-scale (or the onset of the inertial range)~\citep{perez20b}, which is roughly consistent with the lower bound in equation~\eqref{eq:omg_range}.

%\vspace*{8in}

%Table~\ref{tab:results_all} shows

%\newpage

% \begin{figure}
%     \centering
%     \includegraphics{ekperp_spectrum.png}
%     \caption{Reconstructed magnetic energy spectrum vs $k_\perp$}
%     \label{fig:bspectrum}
% \end{figure}
\section{Conclusions\label{sec:conclusions}}
In this work, we presented an analysis of four 24~h intervals during the first four \psp~close encounters to investigate the applicability of Taylor's Hypothesis in the framework of a recent methodology~\citep{bourouaine20}. This new methodology is based on a phenomenological ``sweeping'' model of the spacetime correlation function of MHD turbulence, which was validated against numerical simulations of RMHD turbulence~\citep{perez20b}. As opposed to previous models of the spacetime correlation~\citep{servidio11,lugones16,narita17a}, the BP19 phenomenology suggests that the temporal decorrelation of small scales is entirely due their random advection by velocity fluctuations in the energy-containing range. The BP19 model is broadly applicable to Alfv\'enic solar wind streams, such as those recently observed by \psp~that are believed to originate from a small equatorial coronal hole measured by \psp~\citep{bale19}. The validity of TH in this model depends on a single parameter that measures the ratio between the velocity rms of large-scale fluctuations and the spacecraft speed, perpendicular to the local field, with respect to the plasma frame, $\epsilon=\delta u_0/V_\perp\sqrt 2$. The only assumptions in this methodology are that the turbulence is Alfv\'enic and strong (in the critically balance sense), and that the sampling direction is sufficiently oblique that it can be considered nearly perpendicular to the field. Solar wind observations have been found to be largely consistent with a critically-balanced nonlinear cascade and its associated spectral anisotropy, see for instance~\citep{horbury08,chen11a,papen15}. Under these conditions TH hypothesis is expected to remain as a good approximation as long as $\epsilon\lesssim0.5$~\citep{bourouaine20}.

In our analysis, we found that for the intervals we considered during the first four perihelia, the parameter $\epsilon$ remains below 0.2 at sampling angles greater than $30^\circ$, which can be considered sufficiently oblique. For these values of $\epsilon$, TH is found to hold, irrespective of the value of the Alfv\'en Mach number $M_{\rm A}$. Although TH remains approximately valid, in the sense that frequency spectrum can be interpreted as the one-dimensional spatial energy spectrum with respect to the streamwise direction, the frequency spectrum was used to reconstruct the field-perpendicular energy spectrum $E(k_\perp)$, which measures the spectral energy distribution of the turbulence with respect to the angle-integrated wavenumber $k_\perp$. When TH is used to recover $E(k_\perp)$, a frequency broadening similar to the one obtained in the BP19 phenomenology arises, resulting in an overall decrease of the fluctuation power at each frequency. For the empirical values of $\epsilon$, below 0.2, the broadening is very similar whether TH or the BP19 methodology is used. 

The methodology we presented to reconstruct the spatial energy spectrum from measurements of the frequency spectrum in the spacecraft frame can be applied to measurements from future perihelia closer to the sun, where one expects the value of $\epsilon$ to be larger. This methodology can be summarized in the following straightforward steps:
\begin{enumerate}
    \item The timescale $T=1/f_{\rm b}$ corresponding to the onset of the inertial range of velocity fluctuations is obtained from the spectral break frequency $f_{\rm b}$ that separates the $f^{-1}$ from the inertial range.
    \item Temporal signals for the outer-scale velocity $\vec v'$ and magnetic field $\vec B'$ are obtained via the moving averages defined in equations~\eqref{eq:ma}. Mean and rms values for these signals are obtained, according to equations~\eqref{eq:rms}.
    \item $\theta_{VB}(t)$ signal is calculated from equations~\eqref{eq:thetavb} and used to group \psp~measurements into angular bins of $\Delta\theta=10^\circ$ around angles $\theta_i=10^\circ,20^\circ,\ldots,90^\circ$.  In order to obtain reliable averages, the selected intervals must be long enough to contain a large statistical sample in each angular bin. Figure~\ref{fig:epsilon_e2} shows that the number of samples becomes smaller with increasing $\theta_{VB}$.
    \item The value of the $\epsilon$ parameter as a function of the sampling angle $\theta_{VB}$ is calculated from equation~\eqref{eq:etheta}
    \item Conditioned correlation functions, as defined by equation~\eqref{eq:ctheta}, are calculated. Resulting correlations are used to compute the power spectral density (PSD) via the Fourier transform. A reliable estimate of $C(\tau,\theta_{VB})$ requires a large number of statistical samples at each $\tau$ and that the correlation drops to nearly zero for the largest time lag $\tau$.
    \item The spatial spectrum $E(k_\perp)$ is obtained from equation~\eqref{eq:eperp} for each angle. If the anisotropy assumption is correct and the sampling angle sufficiently oblique, the reconstructed spectrum should be independent of the angle~\citep{papen15,bourouaine20}, as seen in Figure~\ref{fig:pspectra}. The agreement obtained for these four angles becomes better at smaller scales, consistent with Kraichan's sweeping hypothesis.
    \end{enumerate}
The main advantage of the methodology that we present in this work is that allows one to obtain the energy distribution associated with spatial scales in the plasma frame. The spectral indices determined from power-law fits of the measured frequency spectrum accurately represent the spectral indices associated with the underlying spatial spectrum of turbulent fluctuations in the plasma frame. In spite of a small frequency broadening due to large-scale sweeping, the spatial spectrum can still be recovered to obtain the distribution of fluctuation's energy  among scales in the plasma rest frame.

\begin{acknowledgements}
   JCP was partially supported by NASA grants NNX16AH92G, 80NSSC19K0275 and NSF grant AGS-1752827. SB was supported by NASA grants NNX16AH92G, 80NSSC19K0275 and 80NSSC19K1390. CHKC is supported by STFC Ernest Rutherford Fellowship ST/N003748/2 and STFC Consolidated Grant ST/T00018X/1.  Parker Solar Probe was designed, built, and is now operated by the Johns Hopkins Applied Physics Laboratory as part of NASA’s Living with a Star (LWS) program (contract NNN06AA01C). Support from the LWS management and technical team has played a critical role in the success of the Parker Solar Probe mission.
\end{acknowledgements}

\bibliographystyle{aa} % style aa.bst

% \bibliography{MyLibrary.bib} % your references Yourfile.bib

\end{document}